\documentstyle[preprint,aps,psfig]{revtex}
\tightenlines
\begin{document}
\title{ Quarkonia in Hamiltonian Light-Front QCD}
\author{ Martina M. Brisudov\' a, Robert J. Perry and Kenneth G. Wilson\\
{\it Department of Physics}\\ {\it The Ohio State University,
 Columbus, OH 43210.} }
\date{10 July 1996}
\maketitle
\abstract{A constituent parton picture of hadrons with logarithmic confinement
naturally arises in weak coupling light-front QCD.  Confinement provides a
mass gap that allows the constituent picture to emerge.
 The effective renormalized Hamiltonian is computed to
${\cal O}(g^2)$, and used to study charmonium and bottomonium.  Radial
and angular excitations can be used to fix the coupling $\alpha$, the quark
mass $M$, and the cutoff
$\Lambda$.  The resultant hyperfine structure is very close to experiment. } 

\pacs{12.38.Lg,11.10.Gh,14.40.Gx}

The solution of Quantum Chromodynamics in the nonperturbative domain
remains one of the most important and interesting unsolved problems in
physics.  The basic assumption upon which our work is based is
that it is possible to {\it derive a constituent picture for hadrons from
QCD} \cite{thelongpaper,P2,us}.  If this is possible, nonperturbative bound 
state problems in
QCD can be approximated as coupled, few-body Schr{\"o}dinger equations.

To arrive at a constituent approximation to QCD, we first want to separate 
vacuum fluctuations. This is achieved by formulating the theory on the 
light-front where the vacuum
is trivial in presence of cutoffs. 
We start with the canonical light-front Hamiltonian, regulated by a cutoff 
on light-front energy, and use the similarity renormalization group to 
renormalize \cite{similarity}.
We expect the Hmailtonian to contain novel finite counterterms when there 
is spontaneous symmetry breaking \cite{thelongpaper}. The vacuum is 
unchanged by spontaneous breaking.

Renormalization of light-front Hamiltonians is more complicated 
than that of Lagrangians because  
many symmetries are not kinematically manifest, because 
 there is a separate power counting for longitudinal 
and transverse directions, and because locality is violated in the 
longitudinal direction.

The basic idea of the similarity renormalization group is illustrated in Fig.1. 
 Fig. 1(a) schematically shows a regulated bare light-front Hamiltonian matrix. 
The regulator used for illustration 
is a cutoff restricting light-front energies, and it 
makes the Hamiltonian matrix finite. 
There are different ways to impose the cutoff, but the specifics are not 
important for this discussion.
The bare 
Hamiltonian contains couplings between all energy scales. This is the 
source of ultraviolet divergences. In order to renormalize the 
Hamiltonian, 
one needs to find counterterms that remove dependence on the cutoff. 
The similarity renormalization group is based on the following 
observation: if a Hamiltonian has a band diagonal form, as the Hamiltonian 
in fig. 1(b), then no ultraviolet divergence can appear at any finite order of 
perturbation theory because of the finite width of the Hamiltonian, as long as 
its matrix elements are finite. Therefore, if one can find a similarity 
transformation that transforms a bare Hamiltonian $H_B$ as in fig. 1(a) to a 
band diagonal Hamiltonian $H'$ as in fig. 1(b):
\begin{eqnarray}
H'= S H_B S^{\dagger} + \ \ {\rm counterterms},
\end{eqnarray}
it is possible to identify counterterms by requiring that the matrix elements 
of the band diagonal Hamiltonian are independent of the regulator.
This ensures that physical observables will also be independent of the 
regulator.

The similarity renormalization can be done in steps or continuously
 \cite{similarity}. In what follows, we use a discrete 
 perturbative formulation around 
 the free light-front Hamiltonian.

\begin{figure}
%\centerline{(a)}
\begin{eqnarray*}
{\rm (a)}& \left(  \begin{array}{ccccccccc} 
\  \  a_0 \ & \  a \ & \  a \ & \  a \ & \    a \  & \  a \  & ... &\   a
\  & \  a \   \ \\ \
\  a& \  a& \  a& \  a&  \  a& \  a& ... &\   a& \  a \  \\ \ 
\  a& \  a& \  a& \  a&  \  a& \  a& ... &\   a& \  a \  \\ \ 
%\  a& \  a& \  a& \  a& \  a& \  a& ... &\   a& \  a \  \\ \ 
\  a& \  a& \  a& \  a& \  a& \  a& ... &\   a& \  a \  \\ \ 
:& :& :& :& :& :& ::: & :& :\ \\ \ 
%& .& .& .& .& .& ... & .& .\\ \ 
a& a& a& a& a& a& ... & a& a \  \\ \ 
a& a& a& a& a& a& ... & a& a_{\Lambda _0}  \ 
\end{array} \right) &
\\ \\ 
\\ {\rm (b)}
 & \left(  \begin{array}{ccccccccc}
\  \bf A_0 \  & \  \bf A\  & \  0 \  & \  0 \  &  \   0 \  & \   0 \  & \  ...  
\   & \   0 \  & \  0 \  \\
\  \bf A \  & \  \bf A\  & \ \bf  A \  &  \  0 \  & \   0 \  & \   0 \  & \  ...  
\   & \   0 \  & \  0 \  \\
0& \bf A& \bf A& \bf A& 0&  0& ... & 0& 0 \\
0& 0& \bf A& \bf A& \bf A&  0& ... & 0& 0 \\
%0& 0& 0& \bf A& \bf A& \bf A& 0& ... & 0& 0 \\
%0& 0& 0& 0& \bf A& \bf A& \bf A& ... & 0& 0 \\
%0& 0& 0& 0& 0& \bf A& \bf A& ... & 0& 0 \\
:& :& :& :& :& :&  ::: & :& :\\
%.& .& .& .& .&  .& ... & .& .\\
0& 0& 0& 0& 0&  0& ... & \bf A& \bf A \\
0& 0& 0& 0& 0&  0& ... & \bf A& \bf A_{\Lambda _0} 
 \end{array} \right) &
 \end{eqnarray*}
\caption{(a) An example of a bare regulated Hamiltonian 
in light-front energy space. 
Energies run from zero up to an initial cutoff which is indicated by the 
subscripts of the first and the last diagonal elements. $a$ in this figure 
schematically 
indicates nonzero matrix elements, but the matrix elements are not 
necessarily equal. The Hamiltonian couples states of all energy scales.
(b) An example of a  band diagonal Hamiltonian. The energies run up to the 
initial cutoff, but the Hamiltonian couples only states which are close in 
energy. }
\end{figure}

With the divergences removed, the remaining task is to adjust 
finite parts of the counterterms. This is, in principle, achieved by
 restoring Lorentz 
invariance and other exact symmetries
in physical observables. However, if the similarity 
transformation can be done analytically, as  in the 
calculation presented here, it is straightforward to use
 coupling coherence \cite{couplcoh}, which 
uniquely fixes all counterterms without explicit reference to underlying 
symmetries. The basic idea of coupling coherence is the following: 
in the Hamiltonian 
restricted by symmetries, albeit not manifest, the
strengths of all operators are not independent but depend only on a finite
number of independent canonical 
parameters. Under a full renormalization group transformation
(including change of scale and rescaling), the Hamiltonian
reproduces itself in form exactly, apart from the change of the 
explicit cutoff and
the running of  those  few independent couplings.
All dependence on the 
cutoff is absorbed into the  
independent running couplings.
Once one obtains a Hamiltonian that  reproduces itself as the cutoff is
lowered and subtracts the divergences, any initial cutoff can be sent to 
infinity.

The coupling coherent solution at second order in a generic 
interaction $v$ (around the free light-front Hamiltonian $h_0$) 
is \cite{P2}:
\begin{eqnarray}
\lefteqn{ H_{ab} = \langle a|h_0+v|b\rangle  } \nonumber\\
&  -\sum_k v_{ak} v_{kb} \left[
{\theta\bigl(|\Delta_{ak}|-{\lambda^2\over{{\cal P}^+}} \bigr)
\theta\left(|\Delta_{ak}|-|\Delta_{bk}|\right) \over{ -\Delta{ak}} }
+ {\theta\bigl(|\Delta_{bk}|-{\lambda^2\over{{\cal P}^+}}\bigr)
\theta\left(|\Delta_{bk}|-|\Delta_{ak}|\right) \over{ -\Delta{bk}} } 
\right] .
\end{eqnarray}
where  $\Delta{ij}= E_{0i}-E_{0j}$ is the difference in light-front free 
energies, $\lambda^2\over{{\cal P}^+}$ is the similarity scale, and 
sum over states $k$ is limited 
by the initial cutoff and the explicit 
similarity cutoff in eqn. $(2)$. 
A self-energy counterterm is also needed but is not shown (see 
ref. \cite{P2}).

We would like to note that the similarity renormalization scheme which 
results in band diagonal effective Hamiltonians is a renormalization 
method that brings us closer to the desired 
constituent picture of hadrons. Indeed, if the 
constituents are massive, then at any finite order of perturbation theory 
with the band diagonal Hamiltonian, there is only a finite number of Fock 
states that couple to the lowest Fock component in the hadron. 
In our work low energy gluons
 acquire a mass gap due to a mass counterterm
and  due to the confining interaction, and 
high multiplicities of gluons are suppressed. 

After renormalization is completed, one is left with 
the effective Hamiltonian  band diagonal with the width of a hadronic scale.
 The effective Hamiltonian contains complicated potentials, which result from
eliminating the coupling between high and low energy states. It still contains
emission and absorption interactions, but these no longer mix states of high and
low energies. However, if one tries to diagonalize the effective 
Hamiltonian directly, the wave function of a hadron must contain 
arbitrarily many parton components. Instead, we divide the effective 
Hamiltonian into a part $H_0$ which is solved nonperturbatively, and the 
remaining 
part, $V$, is treated in bound state perturbation theory. 
The division is arbitrary, but a choice of $H_0$ missing an important part 
of physics of the system under consideration would lead to divergent bound 
state perturbation theory. Therefore, we want to choose 
 $H_0$  that
approximates the physics relevant for hadronic bound states
 as closely as possible, and at the same time 
we want it to  be  manageable. We take a hint from the constituent quark model
 and include 
constituent masses and two-body potentials produced by the similarity 
transformation.

A major simplification is achieved by not including emission and 
absorption of low energy gluons in $H_0$.
Once the particle-number-changing interactions are put in $V$, 
different Fock states decouple to leading order.
The Hamiltonian $H_0$ provides an approximate $q\bar{q}$ valence quark 
description of mesons. The errors in approximation can be determined from 
bound-state perturbation theory in $V$ which links the $q\bar{q}$ to
 multi-body Fock states.
Mixing of different Fock components first 
enters at 
second order bound state perturbation theory.
The valence approximation 
is best justified for heavy quarkonia. 

Based on the success of the constituent quark model, it is reasonable to choose 
a nonrelativistic limit of the effective Hamiltonian for $H_0$. This 
approximation, too, is best justified for heavy quarkonia.

We have already used the effective Hamiltonian approach to study properties 
of heavy-light mesons, in particular B mesons.
In this letter we present 
numerical results obtained by applying the 
approach to charmonium and bottomonium for which the approximations are 
better justified. We fit 1S, 1P and 2S levels for 
both systems. We then 
predict hyperfine splitting in the charmonium ground state. The prediction 
is in good agreement with experiment.

We find the effective Hamiltonian to ${\cal O}(g^2_{\Lambda} )$. 
The effective Hamiltonian, 
which is generated by the similarity transformation and coupling coherence to 
order $g^2$,
is band-diagonal in light-front energy with respect 
to a hadronic scale 
${\Lambda^2\over{{\cal P}^+}}$,   and it can be written as \cite{us}:
\begin{eqnarray}
H_{\rm eff} = H_{\rm free} + v_1 +  v_{2} +  v_{2 \ {\rm eff}} \  \  ,
\end{eqnarray}
where $H_{\rm free}$ is the light-front kinetic energy (we remind the 
reader that the light-front kinetic energy of a particle with transverse 
momentum ${p}^{\perp}$ and longitudinal momentum $p^+$ is 
${{p^{\perp}}^2 + m^2 \over{p^+}}$), $v_1$ is ${\cal O}(g)$
emission and absorption with
nonzero matrix elements
only between states with energy difference smaller than the
hadronic scale ${\Lambda^2\over{{\cal P}^+}}$.

Let $p_i$, $k_i$ be the light-front three-momenta carried 
by a quark and an antiquark; 
$\sigma _i$, $\lambda_i$ are their light-front helicities; 
$u(p, \sigma )$, $v(k, \lambda )$ are their spinors; %\cite{lebro};
index $i=1,2$ refers to the initial and final states, respectively.
Let $\vec{q} = \vec{p}_1 - \vec{p}_2$ be the exchanged momentum and
$q^- ={ {q^{\perp}}^2\over{q^+}}$. $v_{2} $ in eqn. $(2)$ is
an ${\cal O}(g^2)$
instantaneous interaction with the following matrix element 
for free states 
containing a quark and an antiquark:
\begin{eqnarray}
  & -g_{\Lambda }^2 \bar{u}(p_2, \sigma _2) \gamma ^{\mu} u(p_1, \sigma _1)
\bar{v}(k_2, \lambda_2) \gamma ^{\nu} v(k_1, \lambda _1) 
\langle T_a T_b \rangle
\nonumber\\
 & 
\times  {1\over{{q^+}^2}} \eta_{\mu}\eta_{\nu} \ 
\theta \left({\Lambda^2 \over{{\cal P}^+}} - \vert (p_1^- + k_1^- )
- (p_2^- +k_2^-)\vert \right) \ \ \  ,
\end{eqnarray}
where $\eta_{\mu} v^{\mu} = v^+$ defines the unit vector $\eta_{\mu}$.
 $v_{2 \ {\rm eff}}$ includes
the ${\cal O}(g^2)$ effective
interactions generated by the similarity transformation.
The effective interactions generated to this order contain one-body and
 two-body operators.
In particular, the effective one-body quark operator is:
\begin{eqnarray}
{\alpha_{\Lambda} C_F \over{2\pi P^+}} \left\{
2 {P^+\over{{\cal P}^+}}\Lambda^2 \log\left({P^+\over{\epsilon {\cal P}^+}}
\right)
+ 2 {P^+\over{{\cal P}^+}}\Lambda^2
\log {x_a^2 {P^+\over{{\cal P}^+}}\Lambda^2 \over{x_a{P^+\over{{\cal P}^+}}
\Lambda^2 +M^2  }} \right. \nonumber\\
- {3\over{2}}{P^+\over{{\cal P}^+}}\Lambda^2
+{1\over{2}}{M^2 {P^+\over{{\cal P}^+}}\Lambda^2
\over{x_a{P^+\over{{\cal P}^+}}\Lambda^2 +M^2  }} \nonumber\\
 \left.
+ 3 {M^2\over{x_a}}\log
 {M^2\over{x_a {P^+\over{{\cal P}^+}} \Lambda^2 +M^2  }} \right\} \  \  ,
\end{eqnarray}
where $x_a= {p_a^+\over{P^+}}$ 
is the longitudinal fraction of the momentum carried by the 
constituent under consideration, $M$ is its mass, $P^+$ is the total 
longitudinal momentum of the state, ${\cal P}^+$ is the longitudinal scale 
required in the cutoff by dimensional arguments, and $\epsilon$ is an 
infrared cutoff which is to be taken to zero. The divergence in the 
effective one-body operator exactly cancels against a divergence in the 
effective two-body operator if the state is a color singlet \cite{P2}.

The effective two-body operators have the following matrix elements between 
states containing a quark  $\vec{p}_i$
and an antiquark: 
\begin{eqnarray}
  & -g_{\Lambda }^2 \bar{u}(p_2, \sigma _2) \gamma ^{\mu} u(p_1, \sigma _1)
\bar{v}(k_2, \lambda_2) \gamma ^{\nu} v(k_1, \lambda _1)
\langle T_a T_b \rangle
\nonumber\\
 & \times \left[
 {1\over{q^+}}
D_{\mu \nu}(q)
\left({\theta(\vert D_1\vert-{\Lambda^2 \over{{\cal P}^+}})
\theta(\vert D_1\vert -\vert D_2\vert )\over{D_1}}
+ {\theta(\vert D_2\vert-{\Lambda^2 \over{{\cal P}^+}})
\theta(\vert D_2\vert -\vert D_1\vert )\over{D_2}}\right)
\right] 
\end{eqnarray}
where 
$D_{\mu \nu}(q) = {{q^{\perp}}^2\over{{q^+}^2}}\eta_{\mu}\eta_{\nu}
 + {1\over{q^+}}
\left(\eta_{\mu}{q^{\perp}}_{\nu} + \eta_{\nu}{q^{\perp}}_{\mu}\right)
- g^{\perp}_{\mu \nu}$ is the gluon propagator in light-front gauge,
 $D_1$, $D_2$ are energy denominators:
$D_1 = p_1^- -p_2^- -q^-$ and $D_2 = k_2^- -k_1^- - q^-$.  
It has been shown that $H_{\rm eff}$ contains a logarithmic confining 
interaction in addition to the Coulomb interaction \cite{P2}.

This is the output of the second order similarity transformation for 
$q\bar{q} $ matrix elements.

For the purpose of bound state calculations,
we split the effective Hamiltonian $(4)$
 into $H_0 $, which is solved nonperturbatively, and
 $V=H_{\rm eff}-H_0$.
First, we make a  nonrelativistic  reduction of the effective Hamiltonian
$(4)$. In the nonrelativistic limit, the light-front scale 
${\Lambda^2\over{{\cal P}^+}}$ is naturally replaced by 
${\cal L}\equiv{\Lambda^2\over{{\cal P}^+}}{P^+\over{2M}}$, where 
$M$ is the mass 
of the heavy quark, and ${\cal L}$ carries the dimension of mass \cite{us}.
Further, light-front momenta are naturally replaced by  center-of-mass 
equal-time momenta in the nonrelativistic limit \cite{us}.

The spin-independent part of the 
two-body effective interactions includes a short-range Coulomb potential and 
a rotationally noninvariant
 long-range logarithmic potential. The confining potential arises due to an 
incomplete cancellation  allowing the gluons becoming 
nonperturbative. The confining potential is a complicated function
but in the nonrelativistic limit  it can be double Fourier transformed (for 
longitudinal and transverse separation),  and expanded in even Legendre 
polynomials \cite{us}.

 For $H_0$ we choose the nonrelativistic reduction of 
the kinetic energy, the effective one-body operators, Coulomb potential and 
rotationally symmetric part of the confining potential with constituent 
masses.
The Hamiltonian $H_0$ is:
\begin{eqnarray}
H_0 = 4M\left[ -{1\over{2m}}\vec{\nabla}^2 +\tilde{\Sigma}
- { C_F \alpha \over{r}} +{ C_F \alpha {\cal L} \over{\pi}} V_0({\cal L}r)
\right],
\end{eqnarray}
\noindent where $m$ is the reduced mass, $\vec{r}$ is an equal-time separation 
between the quark and the antiquark,  and $V_0({\cal L}r)$ is the angular 
average of the confining potential generated by the similarity 
transformation. It depends only on the separation of the quarks:
\begin{eqnarray}
V_0({\cal L}r)   & = & 
2 \log {\cal R}-2 Ci({\cal R}) +4 {Si({\cal R})\over{{\cal R}}}
-2 {(1-\cos{\cal R})\over{{\cal R}^2}}+2{\sin{\cal R}\over{{\cal R}}}
-5 +2\gamma  \  \  , 
\end{eqnarray}
where  ${\cal R}\equiv {\cal L}r$ and $\gamma$ is Euler constant.  
 $\tilde{\Sigma }$ contains the finite shift produced by the self-energies 
after subtracting terms needed to make the confining potential vanish at 
the origin:
\begin{eqnarray}
\tilde{\Sigma } &  =  & 
 {\alpha C_F  {\cal L}\over{  \pi}} 
\left[  
\left( 1+{3M\over{2{\cal L}}}\right)
 \log \left( {M\over{ {\cal L}+M}} \right)  
+{1\over{4}}{M\over{ {\cal L}+M}}
+{5\over{4}} 
\right]  .
\end{eqnarray}

The remaining part of the effective Hamiltonian, $V$, contains, among other 
terms, emission and 
absorption of low energy gluons.
Interactions which change particle number enter at 
second order bound-state perturbation theory, which requires solutions
to the nonperturbative three-body bound-state problem. 

$V$ also contains a rotationally noninvariant part of the effective confining 
potential.
 Our choice of $H_0$ does not lead to any first-order corrections 
to S states
due to the rotationally noninvariant
 part of the potential, and for any $l \neq 0$ state it minimizes the 
number 
of terms which give nonzero corrections, thus making the calculations 
easier.

The spin-dependent part of the two-body 
effective operators is included in $V$, and 
it is treated in first-order bound-state perturbation theory. We will 
consider only the spin-spin hyperfine splitting in the ground state, 
because it 
can be calculated using the lowest order effective Hamiltonian
 \cite{billy,c5}.
After a change of the spinor basis \cite{c5}, the
 spin-spin part of the two-body effective interactions is:
\begin{eqnarray}
v_{spin} =4 (2M)^2 \alpha  C_F {\cal L}^3  
\left[{8 \pi\over{3}} \delta^3(\vec{\cal R}) 
+ {2\over{\pi}}f(\vec{\cal R})\right]
{1\over{4 M^2}} 
\vec{\sigma_a} \cdot \vec{\sigma_b} \  \   .
\end{eqnarray}
The function $f(\vec{\cal R})$ is rotationally noninvariant with respect to the 
angle $\theta $. Its angular average is:
\begin{eqnarray}
\langle f ({\cal R}) \rangle = {2\over{3}}
\left[ {\sin{\cal R}\over{{\cal R}^3}} 
- {2(1-\cos{\cal R})\over{{\cal R}^4}} \right] .
\end{eqnarray}
This completes our discussion on $H_0$ and $V$.

We now want to solve  nonperturbatively the eigenvalue problem for $H_0$:
\begin{eqnarray}
H_{0} \vert P \rangle_{\Lambda} = {\cal M}^2 \vert P \rangle_{\Lambda} \  \  , 
\end{eqnarray}
where ${\cal M}^2$ is the invariant mass of the bound state.
We assume that the scale $\Lambda $ is small enough
so that the state is dominated 
by its $q\bar{q}$ component, i.e.,
\begin{eqnarray}
\vert P \rangle _{\Lambda} = \int {d^2\kappa^{\perp} \ dx \over{2(2\pi)^3 
\sqrt{x(1-x)}}}\psi(\kappa^{\perp},x) 
b^{\dagger} d^{\dagger} \vert 0 \rangle \  \  .
\end{eqnarray}
Let the mass of the bound state be
\begin{eqnarray}
{\cal M}^2 = (2M)^2 +4 M E \   \  ,
\end{eqnarray}
which defines $E$.

The eigenvalue problem for the Hamiltonian $H_0$ leads to a dimensionless 
Schr\"odinger equation \cite{us}: 
\begin{eqnarray}
\left[ -{d^2\over{d\vec{{\cal R}}^2}}
+ c \left(
 {1\over{\pi}}V_{\rm conf}(\vec{\cal R}) + V_{\rm coul}(
{\cal R})\right) \right] \psi({\vec{{\cal R}}}) =
 e \psi({\vec{{\cal R}}}) \  \  ,
\end{eqnarray}
where
\begin{eqnarray}
c & \equiv & {2m\alpha C_F \over{{\cal L}}} , 
\\
e & \equiv & {2m (E-\tilde{\Sigma}) \over{{\cal L}^2}} .
\end{eqnarray}

In reference \cite{us} we show the dimensionless eigenvalue $e$ for a few 
lowest lying states for $c$ ranging up to 1.
From the ratio of the splittings between 1S and 1P, and 1P and 2S 
charmonium 
states, we find that $c$ should be around $0.6$. This is the region where 
 neither the Coulomb nor the 
confining potential dominate, and so there 
is not a simple Bohr scaling analysis 
which would assign powers of the coupling to momenta.
For charmonium, 
we find values of the quark mass 
$M_c$, the cutoff $\Lambda$ and $\alpha = {g^2\over{4\pi}}$ so that: (i)
$c=0.6$ (or, equivalently, the ratio of the 1S-1P splitting to the 
1P-2S splitting 
is roughly correct), (ii) the 
mass of the ground state is ${\cal M}_{1S} =3.0$ GeV, 
and, (iii) the mass of the lowest lying P state is ${\cal M}_{1P} =3.5$ GeV. 
Note that  these values are
reasonable approximations given the magnitude of known corrections. 
We obtain $M_c=1.5$ GeV, $\alpha =0.5$ and
 $\Lambda =1.7$ GeV.
Similarly, for bottomonium we  require $c=1.0$, ${\cal M}_{1S} =9.4$ GeV, and 
${\cal M}_{1P} = 9.9$ GeV leading to a bottom quark mass 
$M_b=4.8$ GeV, $\alpha =0.4$, and
$\Lambda = 3.5$ GeV. It is important to note that this coupling need not 
run like $\alpha _{\bar{\rm MS}}$ or $\alpha_{\rm lattice}$.

With the parameters fixed, we can predict the hyperfine splitting in the 
charmonium ground state using eqn. $(11)$.
The function $f(\vec{\cal R})$ is  rotationally noninvariant, but at $c=0.6$ 
the violation of rotational symmetry is negligible. We  predict that the 
splitting between the ground state vector and singlet in charmonium is 
$0.13$ GeV, in reasonable agreement with experiment ($0.118 \pm 0.002$ GeV).

Next, we evaluate corrections due to the rotationally noninvariant part of the 
confining potential. There are first-order corrections to the P state, and 
the S state is corrected in second-order bound state perturbation theory. 
Corrections to the ground state are consistently a few percent even 
for $c$ as small as $0.1$. Corrections to the excited states at $c=0.6$
 and $c=1.0$ are about 10{\%}. 
This is a reasonable starting point because corrections of one higher power of 
$\alpha$ are of this same order.

In conclusion,  the logarithmic confining potential which arises at second 
order is a promising starting point for QCD 
calculations. Corrections to the energy levels due to rotational symmetry 
violating terms in this potential are negligible for the  ground state, and for 
the lowest excited states they are small enough  
that corrections from higher order terms  
may restore rotational symmetry.
This calculation is not intended to compete with phenomenological 
constituent quark model calculations. It is intended as an initial crude 
step toward and accurate first principles, light-front QCD calculation.

Our work was supported by the National Science Foundation under grant 
PHY-9409042.

\end{document}